\documentclass[letterpaper,twocolumn,english,aps,prl,showkeys,showpacs,secnumarabic,citeautoscript,floatfix,superscriptaddress]{revtex4-1}
\usepackage[T1]{fontenc}
\usepackage[latin9]{inputenc}
\usepackage{amsmath}
\usepackage{amssymb}
\usepackage{graphicx}
\usepackage[caption=false]{subfig}
\usepackage{wasysym}
\usepackage{esint}

\makeatletter

\pdfpageheight\paperheight
\pdfpagewidth\paperwidth

\usepackage{color}
\usepackage[usenames,dvipsnames]{xcolor}
\usepackage{hyperref}

\definecolor{refColor}{HTML}{EA00F2}
\definecolor{figColor}{HTML}{008DF2}
\definecolor{urlColor}{HTML}{00AEF2}

\hypersetup{
    colorlinks=true,               
    linkcolor=figColor,            
    citecolor=refColor,            
    filecolor=magenta,             
    urlcolor=urlColor              
,linktoc=page
}

\AtBeginDocument{
  
}

\makeatother

\usepackage{babel}
\begin{document}

\title{A generic theory for Majorana zero modes in 2D superconductors}

\author{Cheung Chan}

\affiliation{International Center for Quantum Materials, School of Physics, Peking
University, Beijing, 100871, China}

\affiliation{Collaborative Innovation Center of Quantum Matter, Beijing 100871,
China}

\author{Lin Zhang}

\affiliation{International Center for Quantum Materials, School of Physics, Peking
University, Beijing, 100871, China}

\affiliation{Collaborative Innovation Center of Quantum Matter, Beijing 100871,
China}

\author{Ting Fung Jeffrey Poon}

\affiliation{International Center for Quantum Materials, School of Physics, Peking
University, Beijing, 100871, China}

\affiliation{Collaborative Innovation Center of Quantum Matter, Beijing 100871,
China}

\author{Ying-Ping He}

\affiliation{International Center for Quantum Materials, School of Physics, Peking
University, Beijing, 100871, China}

\affiliation{Collaborative Innovation Center of Quantum Matter, Beijing 100871,
China}

\author{Yan-Qi Wang}

\affiliation{International Center for Quantum Materials, School of Physics, Peking
University, Beijing, 100871, China}

\affiliation{Collaborative Innovation Center of Quantum Matter, Beijing 100871,
China}

\author{Xiong-Jun Liu}

\thanks{Corresponding author: xiongjunliu@pku.edu.cn}

\affiliation{International Center for Quantum Materials, School of Physics, Peking
University, Beijing, 100871, China}

\affiliation{Collaborative Innovation Center of Quantum Matter, Beijing 100871,
China}
\begin{abstract}
It is well known that non-Abelian Majorana zero modes (MZM) harbor
at vortex cores in a $p_{x}+\text{i}p_{y}$ topological superconductor, which
can be realized in a 2D spin-orbit coupled system with a single Fermi
surface and by proximity coupling to an $s$-wave superconductor.
Here we show that existence of non-Abelian MZMs is unrelated to the
bulk topology of a 2D superconductor, and propose that such exotic
modes can be resulted in much broader range of superconductors, being
topological or trivial. For a generic 2D system with multiple Fermi
surfaces and gapped out by superconducting pairings, we show that
at least a single MZM survives if there are only odd number of Fermi
surfaces of which the corresponding superconducting orders have vortices,
and such MZM is protected by an emergent Chern-Simons invariant, irrespective
of the bulk topology of the superconductor. This result may
enrich new experimental schemes for realizing non-Aelian MZMs.
In particular, we propose a minimal scheme to realize the MZMs in
a 2D superconducting Dirac semimetal with trivial bulk topology, which
can be well achieved based on the recent cold atom experiments.
\end{abstract}

\maketitle
The quest for realization of non-Abelian Majorana zero modes (MZMs), driven by the pursuit of both fundamental physics and their potential application to fault-tolerant topological quantum computation~\cite{Kitaev2003,Nayak2008}, has been spearheaded by the developments
in p-wave superconductors. Early studies predicted that MZMs exist in $\nu=5/2$ fractional quantum Hall state~\cite{Moore-Read}, at the vortex cores in 2D spinless $p_x+\text{i}p_y$ topological superconductors (SCs)~\cite{ReadandGreen2000}, and at ends of a 1D $p$-wave SC \cite{Kitaev2001}. More recently, it has been proposed that the hybrid systems of s-wave SC and spin-orbit (SO) coupled matters with odd number of Fermi surfaces (FSs) can favor effective
p-wave pairing states, bringing the realization of
MZMs to realistic solid state experiments~\cite{Fu2008Majorana,ZhangCW2008PRL,Sato2009MZM,Sau2010Majorana,Lutchyn2010Majorana,Oreg2010Majorana,Franz2015RevModPhys,Wilczek2009,Alicea2012RPP,Franz2013a}. Motivated by these proposals, numerous experimental studies have been performed to observe Majorana induced zero bias conductance anomalies with different heterostructures formed by s-wave SCs and semiconductor nanowires~\cite{Mourik1003,Deng2012,Das2012,Finck2012}, magnetic
chains~\cite{Nadj-Perge2014}, or topological insulators~\cite{FuLiang2008, JiaJin-Feng2015, JiaJin-Feng2016}.

By far the experimental proposals for MZMs are built on the realization of topological SCs. Note that MZMs in SCs are topological defect modes \cite{JeffreyTeo2010,XGWen2012defect,Qi2013defect,Teo2016RevModPhys}, which correspond to nonlocal extrinsic deformations in the Hamiltonian of the topological system. For example, MZMs in the chiral $p_x+ip_y$ SC harbor at vortices which exhibit nonlocal phase windings of the SC order (a global deformation in the original uniform Hamiltonian). This feature tells that the MZMs at vortices are not intrinsic topological excitations, but extrinsic modes of a SC. In this regard, one may conjecture that the existence of MZMs is not uniquely corresponding to the bulk topology of a SC, and there might be much broader range of experimental systems which can host such exotic modes, besides those based on topologically nontrivial SCs.

In this Letter, we show that the existence of MZMs localized in the
vortex cores does not rely on the bulk topology of
a 2D SC, with which we further propose a minimal experimental scheme to realize MZMs in a 2D system. For a generic 2D normal system with $N$ FSs and gapped out by SC pairings.
We show that the existence of the MZMs at the SC vortices is characterized by an emergent
$\mathbb{Z}_{2}$ Chern-Simons invariant $\nu_{3}$:
\begin{align}
\nu_{3} & =\sum^N_{i}n_{i}w_{i}\;\mathrm{mod}\;2,\label{eq:nw}\\
w_{i} & \equiv\frac{1}{2\pi}\oint_{\mathrm{FS}_{i}}\nabla\arg\Delta_{\mathbf{Q}_{i}}(\mathbf{k})\cdot d\bold k,\nonumber
\end{align}
where $\Delta_{\mathbf{Q}_{i}}(\mathbf{k})$ is the SC order projected onto the $i$-th FS and is generically momentum dependent, $w_{i}$ counts the phase winding of $\Delta_{\mathbf{Q}_{i}}(\bold k)$ in the $\bold k$-space around the $i$-th FS loop, and $n_{i}$ denotes the integer vortex winding number (vorticity) attached to $\Delta_{\mathbf{Q}_{i}}\rightarrow\Delta_{\mathbf{Q}_{i}}e^{\i n_i\theta(\bold r)}$. A single MZM is protected when the index $\nu_3=1$, while the bulk of the SC, characterized by Chern number if having no symmetry protection, can be topologically trivial.
We then propose a minimal experimental scheme to realize MZMs based on a 2D superconducting Dirac semimetal whose bulk is topologically trivial. The doped 2D Dirac semimetals are topological metals with strong spin-orbit coupling and possess two FSs which can be fully gapped out by a two-component pairing density wave (PDW) SC order, rendering a trivial 2D superconducting phase with zero Chern number. However, we find that a protected MZM is obtained in the half-vortex regime, namely, only one of two SC order components in the PDW phase is attached with a single vortex, giving a nonzero $\nu_3$ index. The proposed 2D SO coupled Dirac semimetal can be well achieved with the recent cold atom experiments.

\textit{Generic theory}.\textendash We start with the proof of the generic theorem given in Eq.~\eqref{eq:nw} for the Chern-Simons invariant, which governs the existence of the MZM in a 2D (class D) superconductor. For a system with multiple normal bands and FSs, the superconducting pairings may occur within each FS (intra-FS pairings) and between different FSs (inter-FS pairings). As we shall discuss later, the theorem~\eqref{eq:nw} is not affected by inter-FS pairings. Thus for convenience, we consider below the generic SC Hamiltonian with only intra-FS pairings, given by
\begin{equation}
H=\sum_{\mathbf{k}}C_{\mathbf{k}}^{\dagger}\hat{H}_{0}C_{\mathbf{k}} +\sum_{{i},\mathbf{k}}c_{\mathbf{Q}_{i}+\mathbf{k},\alpha}^{\dagger}\hat{\Delta}^{\alpha\beta}_{\mathbf{Q}_{i}} c_{\mathbf{Q}_{i}-\mathbf{k},\beta}^{\dagger}+\mathrm{h.c.},\label{eq:H_generic}
\end{equation}
where $C_{\mathbf{k}}=(c_{\alpha,\mathbf{k}},c_{\beta,\mathbf{k}},\cdots,c_{\gamma,\mathbf{k}},\cdots)^{\text{T}}$,
with $\alpha$ incorporating the band and spin indices, the normal band Hamiltonian $\hat{H}_{0}(\mathbf{k})$
is considered to have $N$ FSs, and the pairing matrix element $\hat{\Delta}^{\alpha\beta}_{\mathbf{Q}_{i}}\propto\langle c_{\mathbf{Q}_{i}/2+\mathbf{k},\alpha}c_{\mathbf{Q}_{i}/2-\mathbf{k},\beta}\rangle$ regarding the $i$-th FS has a central-of-mass momentum $\mathbf{Q}_{i}$. Here for convenience we take that each FS is circular and centered at a
momentum $\mathbf{Q}_{i}/2$. Note that we can always continuously deform the FSs to be circular
without changing topology of the system, as long as the bulk gap keeps open during the deformation. In general the SC order exhibits spatial modulation in the real space, rendering the PDW or Fulde-Ferrell-Larkin-Ovchinnikov state~\cite{Fulde1964,Larkin1965}, and bears the form $\hat\Delta(\mathbf{r})=\sum_{i}\hat\Delta_{\mathbf{Q}_{i}}e^{\text{i}\mathbf{Q}_{i}\cdot\mathbf{r}}$.
Note that each PDW component $\hat\Delta_{\mathbf{Q}_{i}}$ possesses a $U(1)$ symmetry, implying that each of them can be attached with a vortex of winding number $n_{i}$ independently, giving $\hat\Delta(\mathbf{r})=\sum_{i}\hat{\Delta}_{\mathbf{Q}_{i}}e^{-\text{i}n_{i}\theta(\mathbf{r})+\text{i}\mathbf{Q}_{i}\cdot\mathbf{r}}$, with  $\theta(\mathbf{r})$ being the vortex phase profile. Each vortex can host a protected MZM if the Chern-Simons index $\nu_3$ is nontrivial.

To compute the Chern-Simons invariant $\nu_3$ which is defined in 3D space, we parameterize the Bogoliubov de Gennes (BdG) Hamiltonian by taking the phase $\phi\in[0,2\pi)$ of the SC order $\hat{\Delta}_{\mathbf{Q}_{i}}e^{-\text{i}n_{i}\phi}$ as a synthetic dimension of ring geometry $S^1$. Together with the 2D physical space, the bulk BdG Hamiltonian can then be written down in a synthetic 3D torus $T^3=T^2\times S^1$ spanned by $(\mathbf{k}, \phi)$. In the synethetic 3D space, the $\mathbb{Z}_{2}$ Chern-Simons invariant~\cite{Pontryagin,JeffreyTeo2010,WenXiao-Gang2008,SI} can be calculated by
\begin{align}
\nu_{3} & =-\frac{1}{4\pi^{2}}\int_{T^{2}\times S^{1}}\mathcal{Q}_{3}\;\mathrm{mod}\;2\label{eq:nu_3}\\
\mathcal{Q}_{3} & =\mathrm{Tr}\left[\mathcal{A}d\mathcal{A}-\frac{2\text{i}}{3}\mathcal{A}^{3}\right],\nonumber
\end{align}
where the elements of one-form Berry connection are given by $\mathcal{A}_{\lambda\lambda'}(\bold k,\phi)=\text{i}\langle \psi_{\lambda}|\bold d\psi_{\lambda'}\rangle$, with $|\psi_{\lambda}\rangle$ denoting the corresponding eigenvector of the BdG Hamiltonian, and the trace is performed on the filled bands.

A direct computation of the index $\nu_3$ for the generic case is not realistic. To simplify the study we shall take the advantage that the topology of the system is unchanged under any kind of continuous deformation without closing bulk gap. For this we further adiabatically deform the Hamiltonian $H$ to a new form $H'\equiv H[\hat{\Delta}_{\mathbf{Q}_{i}}\rightarrow\hat{\Delta}_{\mathbf{Q}_{i}}\Omega_{\mathbf{Q}_{i}}(\mathbf{k})]$,
where $\Omega_{\mathbf{Q}_{i}}(\mathbf{k})$ is a positive real smooth
truncation function with $\Omega_{\mathbf{Q}_{i}}(\vec{S}_{i})=1$ inside
the orientable vector area $\vec{S}_{i}$ enclosed by the $i$-th FS loop centered at $\mathbf{Q}_{i}$, and decays to zero at a short
distance beyond this area. Since the system remains fully
gapped for the continuous deformation, the invariant $\nu_{3}$
can be evaluated over $H'$. Denoting by $\mathcal{\vec{F}}_{i}$ the vector area with $\Omega_{\mathbf{Q}_{i}}(\mathbf{k})\neq0$, It is straightforward to show that the invariant given in Eq.~\eqref{eq:nu_3} can be reduced to the integral over the disjoint union $\bigsqcup_{i}\vec{\mathcal{F}}_{i}\times S^{1}$~\cite{SI}, which facilitates our further study.

While in general the Hamiltonian $\hat H_0$ incorporates multiple normal bands, we can consider the weak SC pairing regime, in which case only the states around each FS will be effectively paired up. Ignoring the pairing between a state around FS and that from other bands does not affect the topology of the system. In this way, the BdG $H'$ further reduces to an effective one-band form in the eigen-basis $u_{\mathbf{k}}$ of $\hat{H}_{0}$. In particular, for the momentum $\mathbf{k}\in\vec{\mathcal{F}}_{i}$ around a specific FS centered at momentum $\mathbf{Q}_{i}$, the effective BdG Hamiltonian takes the form
\begin{align}
h_{{i}}(\mathbf{k},\phi) =\left[\begin{array}{cc}
\epsilon_{\mathbf{Q}_{i}+\mathbf{k}} & \Delta_{\mathbf{Q}_{i}}(\mathbf{k})\Omega_{\mathbf{Q}_{i}}e^{-\text{i}n_{i}\phi}\\
\Delta_{\mathbf{Q}_{i}}^{*}(\mathbf{k})\Omega_{\mathbf{Q}_{i}}e^{\text{i}n_{i}\phi} & -\epsilon_{\mathbf{Q}_{i}-\mathbf{k}}
\end{array}\right]\label{eq:h_Q}
\end{align}
where $\Delta_{\mathbf{Q}_{i}}(\mathbf{k})\equiv\langle u_{\mathbf{k}}|\hat{\Delta}_{\mathbf{Q}_{i}}|u^{*}_{-\mathbf{k}}\rangle$ is the pairing term projected onto the $i$-th Fermi surface. Note that $\Delta_{\mathbf{Q}_{i}}(\mathbf{k})$ has captured the original band topology.  The eigenstates of  $h_{\mathbf{Q}_{i}}$ take the form $|\psi_{\mathbf{k}\pm}\rangle=(\alpha_{\mathbf{k}\pm}u_{\mathbf{k}},\beta_{\mathbf{k}\pm}u_{-\mathbf{k}}^{*})^{\text{T}}$. Then $\nu_{3}$ can be decomposed into $\nu_{3}=\sum_i\nu_3^{(i)}$
(``mod 2'' temporarily omitted), and
\begin{eqnarray}\label{eq:nu3}
\nu_{3}^{(i)} =-\frac{1}{4\pi^{2}}\int_{\mathcal{\vec{F}}_{i}\times S^{1}}[\mathcal{A}_{\phi}\nabla_{\mathbf{k}}\times\mathcal{A}_{\mathbf{k}}+ \mathcal{A}_{\mathbf{k}}\times\nabla_{\mathbf{k}}\mathcal{A}_{\phi}]d\phi d^{2}\mathbf{k}\nonumber
\end{eqnarray}
for each $\vec{\mathcal{F}_{i}}$, where $\mathcal{A}_{\phi}=\text{i}\langle\psi_{\mathbf{k}-}|\partial_{\phi}|\psi_{\mathbf{k}-}\rangle$,
and $\mathcal{A}_{\mathbf{k}}\equiv(\mathcal{A}_{k_{x}},\mathcal{A}_{k_{y}})=\text{i}\langle\psi_{\mathbf{k}-}|\nabla_{\mathbf{k}}|\psi_{\mathbf{k}-}\rangle$,
with $\nabla_{\mathbf{k}}\equiv(\partial_{k_{x}},\partial_{k_{y}})$.
The above result can be further simplified by taking the limit $\Delta_{\mathbf{Q}}\rightarrow0^{+}$, in which case the gap becomes infinitesimal at the Fermi surface, and the contribution to $\nu_3$ will completely come from the FS states. It can be derived directly on $\vec{\mathcal{F}_{i}}$ that
$\mathcal{A}_{\phi} =- n_i\Theta_{\vec{S}_i}$ and
$\mathcal{A}_{\mathbf{k}} =(1-2\Theta_{\vec{S}_i})\mathcal{A}_{0,\mathbf{k}}+\Theta_{\vec{S}_i}(\nabla_{\mathbf{k}}\arg\Delta_{\mathbf{Q}_i} +\mathbf{A}^i_{d})$,
where $\Theta_{\vec{S}_i}$ is a step
function equal to $1$ within $\vec{S}_i$ and $0$ otherwise, $\mathcal{A}_{0,\mathbf{k}}\equiv\text{i}u_{\mathbf{k}}^{\dagger}\nabla_{\mathbf{k}}u_{\mathbf{k}}$ represents the Berry connection for the normal band, and $\mathbf{A}^i_{d}$ is the \textit{defect gauge field} as a consequence of the multivalueness of $\arg\Delta_{\mathbf{Q}_i}$ \cite{SI,Kleinert2008}. Substituting these results into the formula of $\nu_3$ yields
\begin{align}
\nu_3=\sum_i\frac{n_i}{2\pi}\int_{\vec{\mathcal{F}_i}}\Theta_{\vec{S}_i}\nabla_{\mathbf{k}}\times(\nabla_{\mathbf{k}}\arg\Delta_{\mathbf{Q}_i} +\mathbf{A}^i_{d})\cdot d^{2}\mathbf{k}.
\end{align}
The above result is exactly the one given in Eq.~\eqref{eq:nw} by observing that the curl of gradient of SC phase vanishes, while the contribution from the defect gauge field $\bold A^i_d$ renders the phase winding of SC order in the momentum space around FS loop~\cite{SI}. This completes the proof. The theorem is still valid if the system has dominant pairing between two different FSs, while then the phase winding of the inter-FS pairing has to be computed in both FSs simultaneously, contributing a trivial number to $\nu_3$.

The above result shows that the existence of MZMs at vortex cores is essentially protected not by the bulk topology of the 2D SC, but by an emerging Chern-Simons invariants $\nu_3$, implying that a non-Abelian MZM can exist in a trivial SC. A famous example can be obtained from a Rashba spin-orbit coupled semiconductor with Zeeman splitting and in proximity to a conventional $s$-wave SC~\cite{Sato2009MZM,Sau2010Majorana}. To obtain a chiral topological SC the chemical potential has to lie within the Zeeman gap and cross the bulk band for once. According to the theorem shown here, even the chemical potential is above the Zeeman gap and crosses two FSs, MZMs can in principle be generated if the SC orders in the two FSs are independent and only one of them is attached with vortex.

\begin{figure}
\includegraphics[width=1.0\columnwidth]{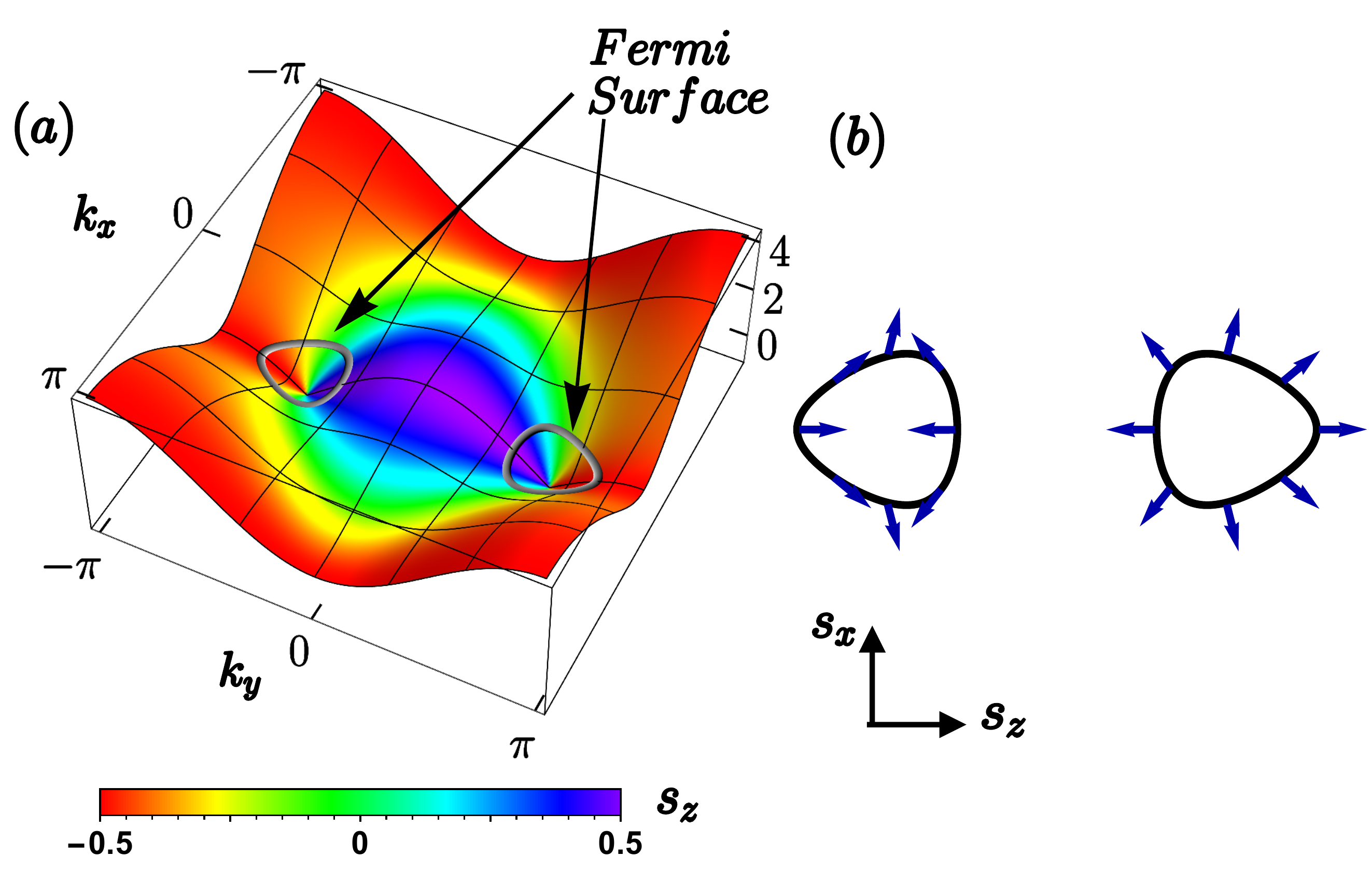}
\caption{\label{fig:band} (a) The band structure of 2D topological Dirac metal
with two Dirac points located at $\mathbf{Q}_{\pm}$, the gray thick
loops around two Dirac points represent the Fermi surfaces, and the
color represents the average value of the spin component $\left\langle s_{z}\right\rangle $.
(b) Schematic of the spin orientations, shown by blue arrows, at
the Fermi surfaces around the Dirac points. Parameters: $t_{x,y}=t_{so}=m_{z}=1$
and the corresponding Dirac node momenta $\mathbf{Q}_{+}=-\mathbf{Q}_{-}=(0,2\pi/3)$.}
\end{figure}

\textit{2D Dirac metal}.\textendash
The theorem in~\eqref{eq:nw} suggests that MZMs can exist in broader range of physical systems. In the following we propose a minimal scheme, which can be readily achieved
based on a recent cold atom experiment~\cite{Wu2016PKU,LiuX-J2017}, for the realization of MZMs. The Hamiltonian takes the form
\begin{eqnarray}
H_{0} & = & \sum_{\mathbf{k}}(c_{\mathbf{k}\uparrow}^{\dagger},c_{\mathbf{k}\downarrow}^{\dagger})\mathcal{H}_{0}\left(\begin{array}{c}
c_{\mathbf{k}\uparrow}\\
c_{\mathbf{k}\downarrow}
\end{array}\right)\label{eq:H_0}\\
\mathcal{H}_{0} & = & (m_{z}-2t_{x}\cos k_{x}-2t_{y}\cos k_{y})\sigma_{z}+2t_{so}\sin k_{x}\sigma_{x}-\mu,\nonumber
\end{eqnarray}
where $c_{\mathbf{k}s}$ ($c_{\mathbf{k}s}^{\dagger}$) is annihilation (creation) operator
with spin $s=\uparrow,\downarrow$, $t_{x,y}$ is the spin-conserved hopping
along $x$/$y$ direction, $t_{so}$ is the spin-flip
hopping amplitude, $m_{z}$ denotes the effective Zeeman coupling, and $\mu$
is the chemical potential. The Hamiltonian in Eq.~\eqref{eq:H_0} describes
a topological Dirac metal for $|m_{z}|<2(t_{x}+t_{y})$, with two
Dirac points at $\mathbf{Q}_{\pm}=(0,\pm\cos^{-1}((m_{z}-2t_{y})/2t_{x}))$
and possesses non-trivial spin texture on the Fermi surfaces (Fig.~\ref{fig:band}).
Note that here the 2D Dirac metal is driven by spin-orbit interaction,
and is distinct from graphene, of which the Dirac points are protected
by symmetry only if spin-orbit coupling is absent~\cite{CharlesL2015}.

\textit{Superconducting phase diagram \& MZM}.\textendash With
the above model, the superfluid (superconductor) states
can be studied by considering an attractive Hubbard interaction. The
total Hamiltonian is $H=H_{0}-U\sum_{i}n_{i\uparrow}n_{i\downarrow}$
for $U>0$. For the multiple Fermi surfaces corresponding to various
Dirac cones, generically one shall consider both the inter-cone (BCS)
and the intra-cone (PDW) pairing orders, described by $\Delta_{2q}=(U/N)\sum_{\mathbf{k}}\left\langle c_{\mathbf{q}+\mathbf{k}\uparrow}c_{\mathbf{q}-\mathbf{k}\downarrow}\right\rangle $,
with $\mathbf{q}=\mathbf{Q}_{\pm}$ or 0 \cite{Moore2012,LiuX-J2016,LiuX-J2017,PhysRevLett.114.237001}
and $N$ is total number of lattice sites. Generally, the order parameter
in real space takes the form
\[
\Delta(\mathbf{r})=\Delta_{0}+\Delta_{2\mathbf{Q}_{+}}e^{2\text{i}\mathbf{Q}_{+}\cdot \mathbf{r}}+\Delta_{2\mathbf{Q}_{-}}e^{2\text{i}\mathbf{Q}_{-}\cdot \mathbf{r}}
\]
and the BCS and PDW orders may compete with each other. Owing to the different
spin-momentum lock at the Fermi surfaces of the two Dirac cones [Fig.~\ref{fig:band} (b)], the inter-cone
BCS pairing cannot fully gap out the bulk spectrum, and leaves four
nodal points. On the other hand, the intra-cone PDW order can fully gap the bulk
in the expense of reducing the translation symmetry. The two types of orders may compete to dominate in different parameter regimes.

\begin{figure}
\includegraphics[width=0.95\columnwidth]{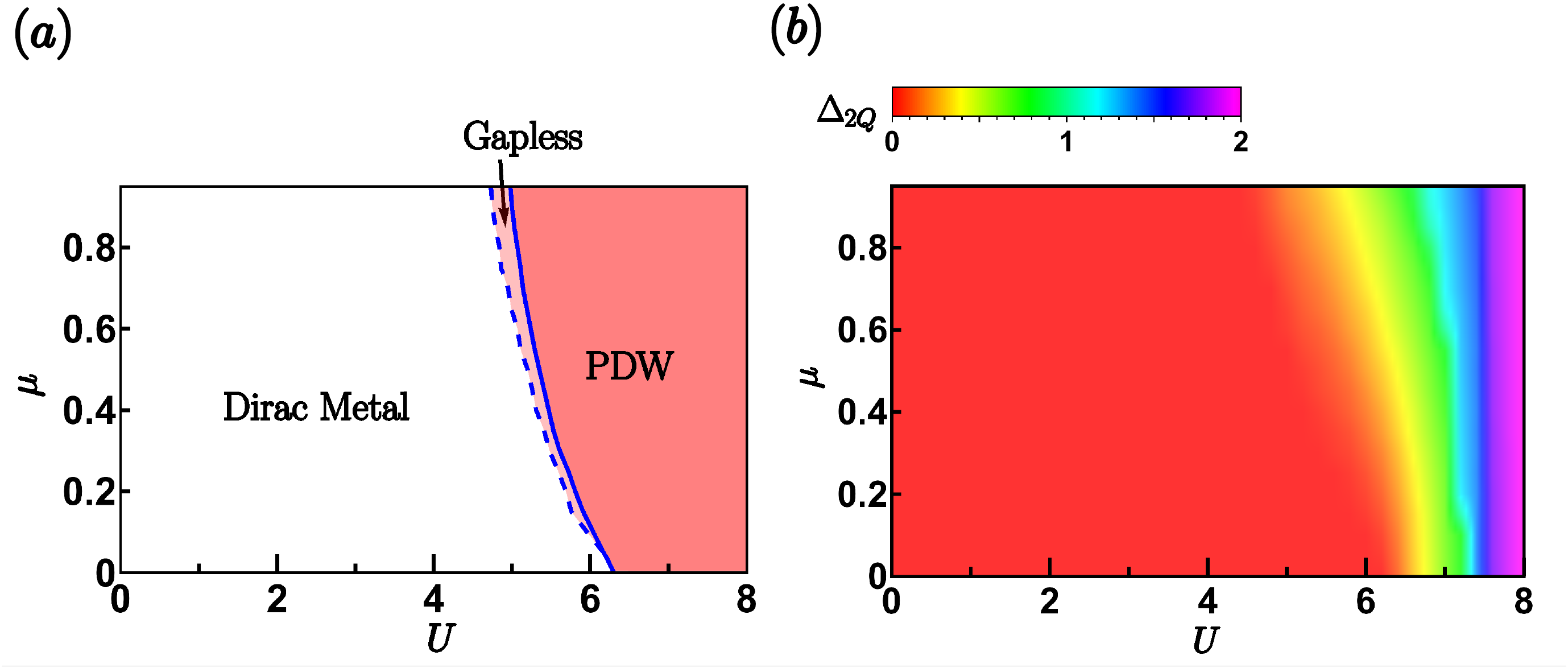}

\caption{\label{fig:phsDiag} (a) Mean field phase diagram of the Dirac-Hubbard
Hamiltonian versus attractive Hubbard interaction $U$ and chemical
potential $\mu$. The BCS order is always suppressed and $\Delta_{0}=0$. In the "Dirac
metal'' phase, all $\Delta_{\mathbf{q}}=0$; in the narrow "Gapless" region, $\Delta_{2\mathbf{Q}_{\pm}}$
are finite but not strong enough to fully gap the system. In the "PDW" phase, the system is fully gapped; (b) Magnitude of PDW order
$\Delta_{2\mathbf{Q}_{\pm}}$. The parameters for numerics are the same as those in Fig.\ \ref{fig:band}.}
\end{figure}

\begin{figure}
\includegraphics[width=1.0\columnwidth]{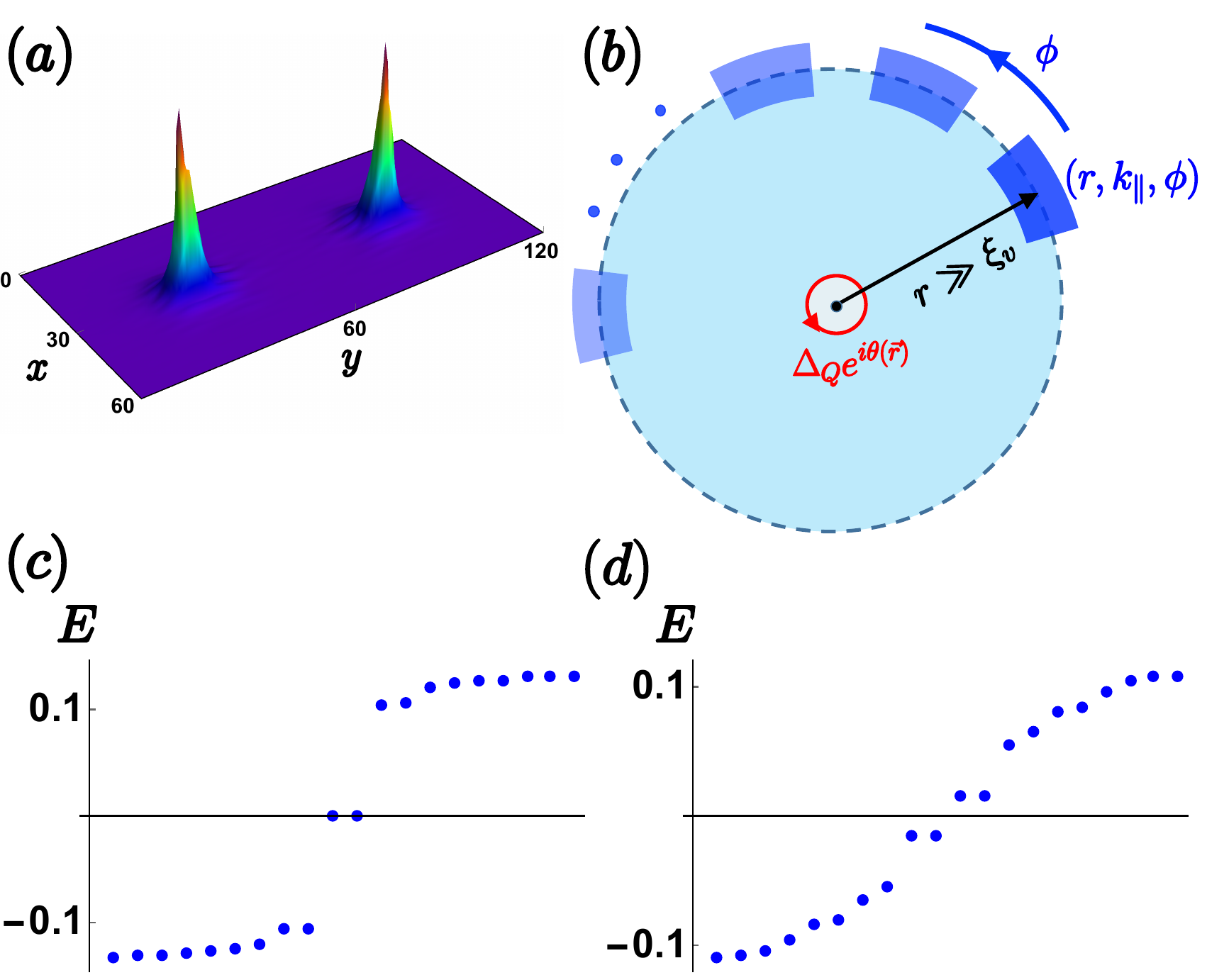}

\caption{\label{fig:MZM} (a) Wave function density $\sum_{s=\uparrow,\downarrow}|\psi_{s}(\mathbf{r})|^{2}$ for MZMs
computed in the Dirac-Hubbard model, with $\mu=0.8$, $U=5.5$,
which gives self-consistent pairing $\Delta_{2\mathbf{Q}_{\pm}}=0.23$
(corresponding to a SC gap of $\Delta_{\mathrm{gap}}=0.28$) and a trivial 2D bulk. System size: $N_{x}=N_{y}/2=60$. The vortices
with opposite unit vorticities are located at $(30,30)$ and $(30,90)$
and the vortex field $e^{\text{i}\theta(\mathbf{r})}$ is attached only to $\Delta_{2\mathbf{Q}_{+}}$. (b) Schematic of the physical origin of MZMs at vortex cores. The vortex core can be viewed approximately as the open boundary of $r$-dimension in the 3D space spanned by $(r, k_{\parallel},\phi)$. Energy spectrum for the half vortex (c) and full vortex (d) regime. The MZMs are obtained in the former case. Other parameter condition is the same as that in Fig.\ \ref{fig:band}.}
\end{figure}

The phase diagram are obtained by self-consistent calculation with proper parameters so that the Dirac points are located at $\bold Q_{\pm}=(0,\pm2\pi/3)$ (see more details in the Supplementary Material~\cite{SI}), as shown in Fig.~\ref{fig:phsDiag}.
It can be found that the BCS pairing is generically suppressed and only the PDW phase
with equal magnitude of $|\Delta_{2\mathbf{Q}_{\pm}}|$ exists. With increasing chemical potential,
the Dirac cone becomes less isotropic (Fig.~\ref{fig:band}])and the Fermi surfaces are less
well-nested. As a consequence, a narrow gapless region with nonzero PDW orders $|\Delta_{2\mathbf{Q}_{\pm}}|\neq0$ is obtained for $\mu>0.05$ [Fig.~\ref{fig:phsDiag}(a)], while the spectrum becomes fully gapped when $|\Delta_{2\mathbf{Q}_{\pm}}|$ increases exceeding some finite value.
In the fully gapped region, one can readily check that the Chern number vanishes for the present (class D) bulk  superconductor~\cite{LiuX-J2017,LIU2014PRL}.

Given the topologically trivial superconducting state here, the system
can still host non-trivial MZM bound to vortices and protected by the Chern-Simons invariant shown above. In general, the vortices proliferated to the PDW order can be $\Delta(\mathbf{r})=\Delta_{2\mathbf{Q}_{+}}e^{2\text{i}\mathbf{Q}_{+}\cdot \mathbf{r}+in_+\theta(\bold r)}+\Delta_{2\mathbf{Q}_{-}}e^{2\text{i}\mathbf{Q}_{-}\cdot \mathbf{r}+in_-\theta(\bold r)}=2\Delta_{2\mathbf{Q}_{\pm}}e^{i(n_++n_-)\theta(\bold r)/2}\cos[2\mathbf{Q}_{+}\cdot \mathbf{r}+(n_+-n_-)\theta(\bold r)/2]$. The minimal regime corresponds to the half-vortex configuration, given by $n_++n_-=\pm1$, while a full vortex is given by $n_++n_-=\pm2$. In particular, in Fig.~\ref{fig:MZM} (a,c) we consider the half vortex regime with two unit vortices of opposite vorticities $\pm2\pi$ (i.e. $n_+=\pm1$) attached only to $\Delta_{2\mathbf{Q}_{+}}$
and located with a finite distance between other in the real space. The real space BdG Hamiltonian with vortices is then numerically
solved and the two lowest energy modes with finite-size energies $E=\pm1.039\times10^{-4}$
are obtained {[}Fig.\ \ref{fig:MZM}(c){]}. Spatial wave function
density $\sum_{s=\uparrow,\downarrow}|\psi_{s}(\mathbf{r})|^{2}$ for one of
the solutions (the other is the same) is plotted in Fig.~\ref{fig:MZM}(a), showing that it is in the zero angular-momentum channel
and well-localized at the vortex cores, thus being a MZM. The physical origin of the exsistence of MZMs can viewed as a direct consequence of \textit{bulk-boundary correspondence}, as illustrated in Fig.~\ref{fig:MZM}(b). Consider the region far away enough form the vortex core so that at each azimuthal angle $\phi$ we can find a microscopic large region with approximately constant SC phase $\theta$. This region can be thought of as a 2D system in $(r, k_{\parallel};\phi)$ with fixed $\phi=\theta$, periodic boundary along $k_{\parallel}$ direction and open boundary along $r$ direction. Combining all such 2D systems with $\phi \in [0,2\pi)$ yields an effective 3D space with periodic boundary with respect to $k_{\parallel}$ and $\phi$, while open boundary along $r$ axis due to the existence of vortex. With this picture when the parameterized 3D system has a nontrivial Chern-Simons invariant $\nu_3$, which is the case for half-vortex regime based on a direct numerical check, MZM is obtained as a boundary zero mode at the vortex core.
In comparison, we have performed a similar calculation by attaching a full vortex with $n_++n_-=2$ to $\Delta_{2\mathbf{Q}_{\pm}}$, which gives a null $\nu_{3}$. In Fig.~\ref{fig:MZM}(d), the corresponding low energy spectrum
reveals that no zero mode but finite energy Andreev bound states are present in the system, consistent
with the $\nu_{3}$ result.

\par
In conclusion, we have developed a generic theory for MZM modes at the vortex cores in 2D superconductors. Our results show that the MZMs are generically protected by an emerging Chern-Simons invariant which can be nontrivial even the bulk of the superconductor is topologically trivial. The result that existence of MZMs is unrelated to the superconducting bulk topology enriches broader range of experimental systems to host non-Abelian MZMs, in particular for the Dirac materials which have even number of Fermi surfaces. As a minimal experimental scheme based on a trivial superconductor/superfluid, we have proposed to realize non-Abelian MZMs with a SO coupled 2D Dirac semimetal which can be fulled gapped out by PDW pairing orders. The protected MZMs have been shown to exist in the half-vortex regime. Such a Dirac semimetal system can be readily realized with the recent cold atom experiment~\cite{Wu2016PKU,LiuX-J2017}. While in the present study we have focused on the MZMs without symmetry protection, it is of great interests to generalize the present theory to the superconductors with protection by symmetry, like time reversal, mirror, or other symmetries.
\par

{\it Note added:} In completing the present manuscript, we are informed of another interesting work by Z. Yan etal, which presents a different model for the realization of MZMs in a trivial 2D superconductor~\cite{ZhongWang2017}.

This work is supported by MOST (Grant No. 2016YFA0301604), NSFC (No.
11574008), and Thousand-Young-Talent Program of China.

\bibliographystyle{apsrev4-1}
\bibliography{ref}

\onecolumngrid

\renewcommand{\thesection}{S-\arabic{section}}
\setcounter{section}{0}  
\renewcommand{\theequation}{S\arabic{equation}}
\setcounter{equation}{0}  
\renewcommand{\thefigure}{S\arabic{figure}}
\setcounter{figure}{0}  

\indent

\begin{center}\large
\textbf{Supplementary Material:\\ A generic theory for Majorana zero modes in 2D superconductors}
\end{center}
In this Supplementary Material we provide the details on the selfconsistent
mean field study and the proof of the Chern-Simons invariant.

\section{Details for the mean field self consistent calculations}

Consider superconductor (SC) order parameters $\Delta_{2\mathbf{q}}=\frac{U}{N}\sum_{\mathbf{k}}\left\langle c_{\mathbf{\mathbf{q}+\mathbf{k}\uparrow}}c_{\mathbf{Q}-\mathbf{k}\downarrow}\right\rangle $,
with $\mathbf{q}=0$ for BCS and $\mathbf{q}=\mathbf{Q}_{\pm}$ for
PDW orders. With the PDW orders, the original Brillouin zone (BZ)
will be folded up into sub-BZ. In the present study, we choose $\mathbf{Q_{+}=-Q_{-}}=(0,2\pi/3)$,
so the folded BZ is $1/3$ of the original BZ, and the mean field
Hamiltonian can be written as
\[
H_{\text{MF}}=\frac{1}{2}\sum_{\mathbf{k}}\Psi_{\mathbf{k}}^{\dagger}\mathcal{H}_{\text{MF}}(\mathbf{k})\Psi_{\mathbf{k}},
\]
\begin{equation}
\mathcal{H}_{\text{MF}}(\mathbf{k})=\left[\begin{array}{cc}
\hat{\mathcal{H}}_{0}(\mathbf{k}) & \hat{\Delta}(\mathbf{k})\\
\hat{\Delta}^{\dagger}(\mathbf{k}) & -\hat{\mathcal{H}}_{0}^{\text{T}}(-\mathbf{k})
\end{array}\right],\label{eq:mf}
\end{equation}
where the basis for the folded BZ is denoted as
\[
\Psi_{\mathbf{k}}=(c_{\mathbf{Q}_{+}+\mathbf{k}\uparrow},c_{\mathbf{k}\uparrow},c_{\mathbf{Q}_{-}+\mathbf{k}\uparrow},(\uparrow\to\downarrow);c_{\mathbf{Q}_{+}-\mathbf{k}\uparrow}^{\dagger},c_{-\mathbf{k}\uparrow}^{\dagger},c_{\mathbf{Q}_{-}-\mathbf{k}\uparrow}^{\dagger},(\uparrow\to\downarrow))^{\text{T}},
\]
with $k_{x}\in[-\pi,\pi)$ and $k_{y}\in[-\pi/3,\pi/3)$. The explicit
form of $\hat{\mathcal{H}}_{0}(\mathbf{k})$ is obtained by restricting
the momentum of the 2D topological Dirac metal Hamiltonian $\mathcal{H}_{0}(\mathbf{k})=(m_{z}-2t_{x}\cos k_{x}-2t_{y}\cos k_{y})\sigma_{z}+2t_{\text{so}}\sin k_{x}\sigma_{x}-\mu$
within a sub-BZ. The order parameter $\hat{\Delta}$ in the matrix
form reads
\begin{equation}
\hat{\Delta}=\left[\begin{array}{cc}
 & \Delta_{[\mathbf{\mathbf{Q}}]}\\
-\Delta_{[\mathbf{\mathbf{Q}}]}
\end{array}\right]\label{eq:delta}
\end{equation}
with
\[
\Delta_{[\mathbf{Q}]}=\left[\begin{array}{ccc}
\Delta_{2\mathbf{Q}_{+}} & \Delta_{2\mathbf{Q}_{-}} & \Delta_{0}\\
\Delta_{2\mathbf{Q}_{-}} & \Delta_{0} & \Delta_{2\mathbf{Q}_{+}}\\
\Delta_{0} & \Delta_{2\mathbf{Q}_{+}} & \Delta_{2\mathbf{Q}_{-}}
\end{array}\right].
\]
Utilizing Eq. \eqref{eq:mf} and Eq. \eqref{eq:delta}, one can interatively
solve the Hamiltonian and compute $\Delta_{2\mathbf{q}}$'s until
convergence. The mean field phase diagram versus attractive interaction
strength $U$ and chemical potential $\mu$ are shown in Fig. 2 in
the main text.

\section{Derivations for the generic reduced formula of the Chern-Simons invariant}

Now we are going to prove that, for a 2D superconductor with vortices,
the Chern-Simons invariant $\nu_{3}$~\cite{Pontryagin,WenXiao-Gang2008,JeffreyTeo2010} 
defined in the base space $(k_{x},k_{y},\phi)\in T^{2}\times S^{1}$,
with $\phi$ denoting the emergent dimension for vorticity,
\begin{align*}
\nu_{3} & =-\frac{1}{4\pi^{2}}\int_{T^{2}\times S^{1}}\mathcal{Q}_{3}\;\mathrm{mod}\;2,\\
\mathcal{Q}_{3} & =\mathrm{Tr}\left[\mathcal{A}d\mathcal{A}-\frac{2\text{i}}{3}\mathcal{A}^{3}\right],
\end{align*}
where $\mathcal{A}_{\lambda\lambda'}(\mathbf{k},\phi)=\text{i}\langle \psi_{\lambda}|d\psi_{\lambda'}\rangle$ is the
one-form Berry connection ($|\psi_{\lambda}\rangle$ is the corresponding eigenvector
of the Hamiltonian, and the trace is performed on the filled bands), takes the following simple form
\[
\nu_{3}=\frac{1}{2\pi}\sum_{i}n_{i}\oint_{\partial\vec{S}_{i}}\nabla_{\mathbf{k}}\arg\Delta_{\mathbf{Q}_{i}}(\mathbf{k})\cdot d\mathbf{k}\;\mathrm{mod}\;2,
\]
where the orientable area $\vec{S}_{i}$ denotes the region enclosed
by the $i$-th Fermi surface ($\partial S_{i}=\text{FS}_{i}$ in the
main text), $\Delta_{\mathbf{Q}_{i}}$ is the superconductor
order parameter projected onto the $i$-th Fermi surface, and $n_{i}$ denotes the integer vortex winding number attached to $\Delta_{\mathbf{Q}_{i}}$.

For a general BdG Hamiltonian
\[
H=\sum_{\mathbf{k}\in T^{2}}C_{\mathbf{k}}^{\dagger}\hat{H}_{0}(\mathbf{k})C_{\mathbf{k}}+\sum_{i,\mathbf{k}}c_{\mathbf{Q}_{i}+\mathbf{k},\alpha}^{\dagger}\hat{\Delta}^{\alpha\beta}_{\mathbf{Q}_{i}}c_{\mathbf{Q}_{i}-\mathbf{k},\beta}^{\dagger}+\mathrm{h.c.},
\]
where $C_{\mathbf{k}}=(c_{\alpha,\mathbf{k}},c_{\beta,\mathbf{k}},\cdots,c_{\gamma,\mathbf{k}},\cdots)^{\text{T}}$,
with $\alpha$ incorporating the band and spin indices. Suppose that the normal band Hamiltonian
$\hat{H}_{0}(\mathbf{k})$ has multiple bands $\epsilon_{n}(\mathbf{k})$,
and only one of such bands, with (normalized) eigenvector $u_{\mathbf{k}}$,
cuts the chemical potential, and we call this band middle band. The
middle band gives rise to $N$ Fermi surfaces with possible
Berry phases, and $\hat{\Delta}_{\mathbf{Q}_{i}}$ are pairing terms that
can fully gap the whole system. The following results can be easily
generalized to the system with multiple middle bands.

We assume that each Fermi surface is circular and centered at some
momentum $\mathbf{Q}_{i}/2$, otherwise one can always continuously
deform the original Hamiltonian to the current form without gap closing.
One can imagine that each Fermi surface is equipped with a PDW order
parameter $\hat{\Delta}_{\mathbf{Q}_{i}}$ and we further assume that
\emph{the system is fully gapped }\textit{only if}\emph{ all the $\hat{\Delta}_{\mathbf{Q}_{i}}$'s
are non-vanishing}. Here we consider some of $\hat{\Delta}_{\mathbf{Q}_{i}}$'s
acquire a winding $\hat{\Delta}_{\mathbf{Q}_{i}}\rightarrow\hat{\Delta}_{\mathbf{Q}_{i}}e^{-\text{i}n_{i}\phi}$
with $\phi\in[0,2\pi)$ and $n_{i}\in\mathbb{Z}$. Together with the 2D physical space, the bulk BdG Hamiltonian can then be written down in a synthetic 3D torus $T^{3}=T^{2}\times S^{1}$ spanned by $(\mathbf{k},\phi)$.

\begin{figure}[!ht]
\subfloat[]{\includegraphics[scale=0.7]{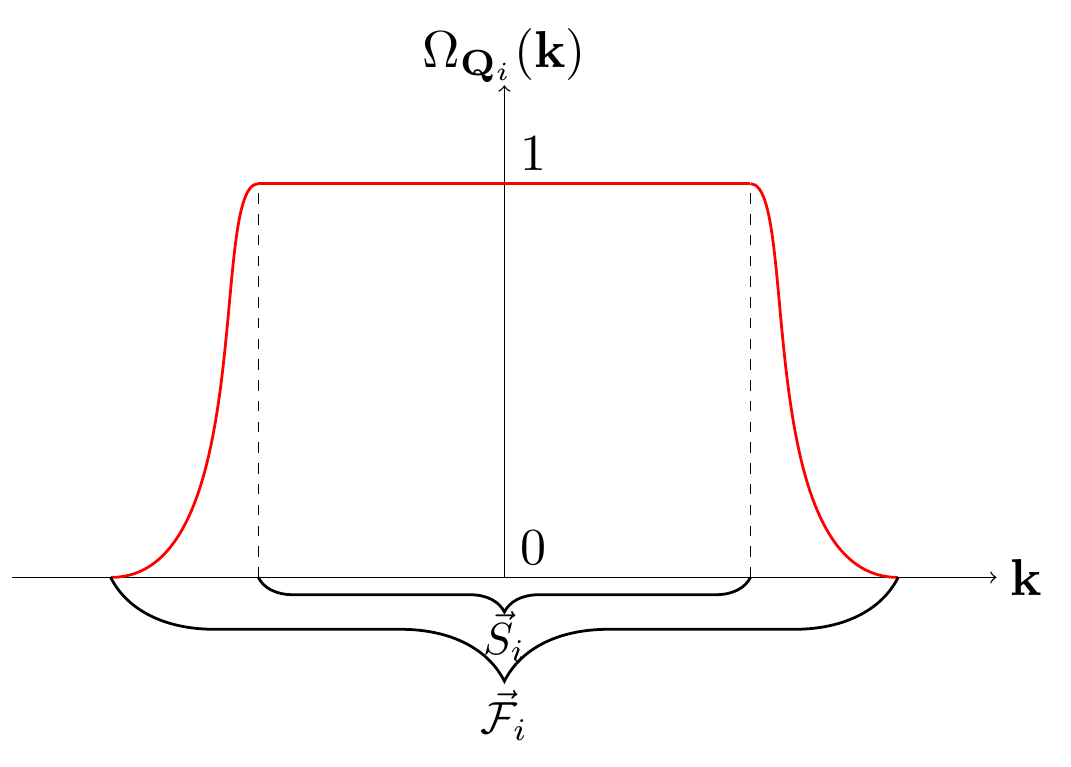}}
\subfloat[]{\includegraphics[scale=0.5]{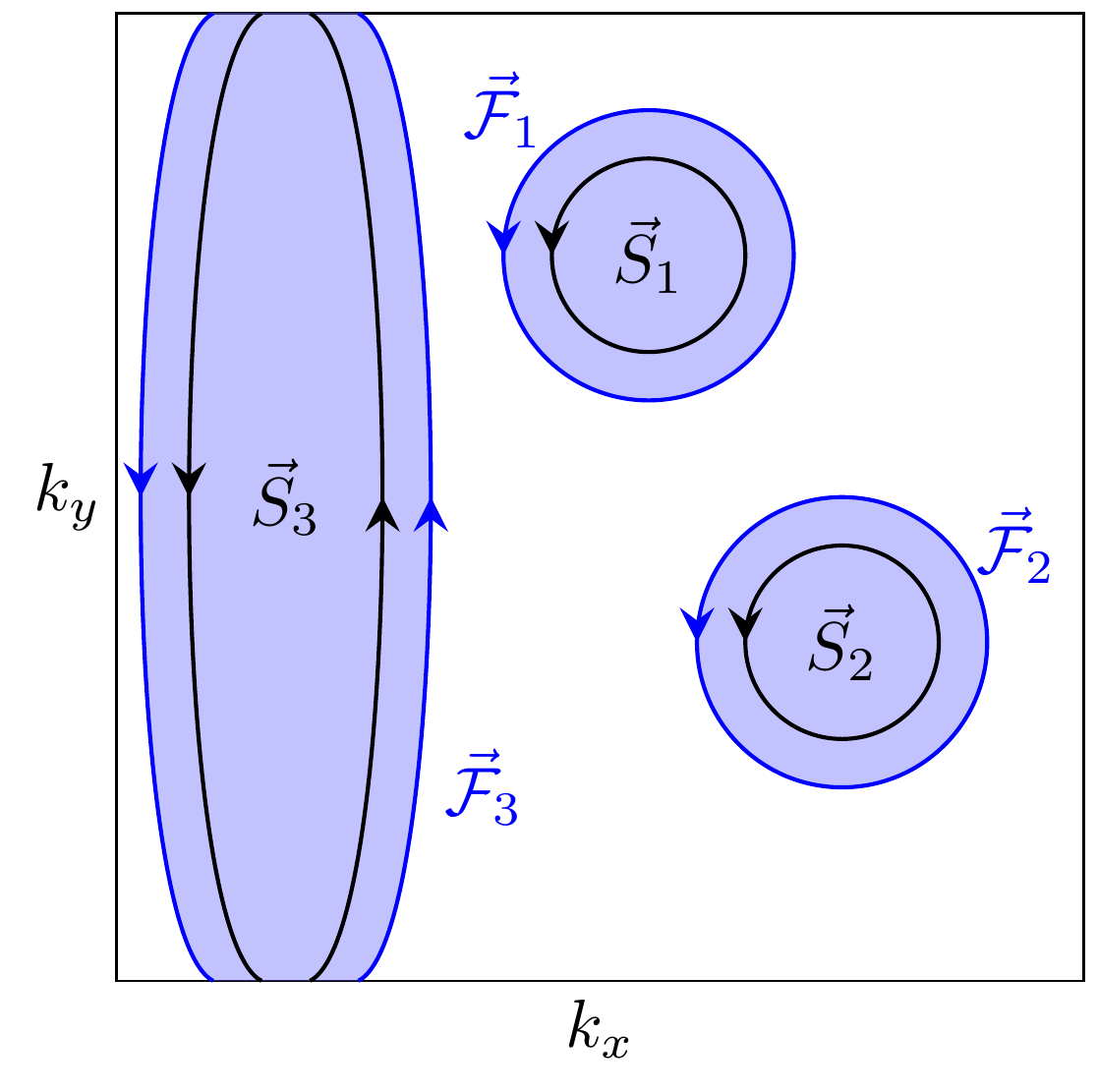}}

\caption{\label{fig:Schematic} (a) Schematic diagram of the real smooth function
$\Omega_{\mathbf{Q}_{i}}(\mathbf{k})$. (b) Schematic diagram of $\vec{S}_{i}$ and $\mathcal{\vec{F}}_{i}$, the black lines denote the Fermi surfaces, the region enclosed by black line denotes $\vec{S}$, and the green area denotes the patch $\mathcal{\vec{F}}$.}
\end{figure}
Consider a continuous deformation $H\rightarrow H'$, with $H'\equiv H[\hat{\Delta}_{\mathbf{Q}_{i}}\rightarrow\hat{\Delta}_{\mathbf{Q}_{i}}\Omega_{\mathbf{Q}_{i}}(\mathbf{k})]$,
where $\Omega_{\mathbf{Q}_{i}}(\mathbf{k})$ is a positive real smooth
truncation function with $\Omega_{\mathbf{Q}_{i}}(\mathbf{k}\in\vec{S}_{i})=1$,
and decays to zero at a short distance from the Fermi surface, as
shown in Fig.\ \ref{fig:Schematic}(a). We denote the region with
$\Omega_{\mathbf{Q}_{i}}(\mathbf{k})\neq0$ as $\vec{\mathcal{F}}=\bigsqcup_{i}\mathcal{\vec{F}}_{i}$
(disjoint union of $\mathcal{N}$ orientable areas $\mathcal{\vec{F}}_{i}$)
and $\vec{\bar{\mathcal{F}}}=T^{2}-\mathcal{\vec{F}}$, note that
$\vec{S}_{i}\subset\mathcal{\vec{F}}_{i}$, as shown in Fig.\ \ref{fig:Schematic}(b).
Since the whole system remains fully gapped for a continuous deformation
$H\rightarrow H'$, we have (``mod 2'' temporarily omitted)
\[
\nu_{3}=-\frac{1}{4\pi^{2}}\int_{T^{2}\times S^{1}}\mathcal{Q}_{3}[H]=-\frac{1}{4\pi^{2}}\int_{T^{2}\times S^{1}}\mathcal{Q}_{3}[H'].
\]

One can easily see that
\[
\int_{\vec{\bar{\mathcal{F}}}\times S^{1}}\mathcal{Q}_{3}[H']=0,
\]
since in the region $\vec{\bar{\mathcal{F}}}$, $\mathcal{A}_{\phi}=0$
and $\partial_{\phi}\mathcal{A}_{k_{x},k_{y}}=0$, then $\mathcal{A}d\mathcal{A}=\text{Tr}[\mathcal{A}_{k_{x}}(\partial_{k_{y}}\mathcal{A}_{\phi}-\partial_{\phi}\mathcal{A}_{k_{y}})+\mathcal{A}_{k_{y}}(\partial_{\phi}\mathcal{A}_{k_{x}}-\partial_{k_{x}}\mathcal{A}_{\phi})+\mathcal{A}_{\phi}(\partial_{k_{x}}\mathcal{A}_{k_{y}}-\partial_{k_{y}}\mathcal{A}_{k_{x}})]dk_{x}\wedge dk_{y}\wedge d\phi=0$,
also $\mathcal{A}^{3}\sim\mathrm{Tr[}\mathcal{A}_{k_{x}}\mathcal{A}_{k_{y}}\mathcal{A}_{\phi}-\mathcal{A}_{k_{y}}\mathcal{A}_{k_{x}}\mathcal{A}_{\phi}]dk_{x}\wedge dk_{y}\wedge d\phi=0$,
hence,
\[
\nu_{3}=-\frac{1}{4\pi^{2}}\int_{\mathcal{\vec{F}}\times S^{1}}\mathcal{Q}_{3}[H']=-\frac{1}{4\pi^{2}}\sum_{i}\int_{\mathcal{\vec{F}}_{i}\times S^{1}}\mathcal{Q}_{3}[H'].
\]

While in general the Hamiltonian $\hat{H}_{0}$ incorporates multiple normal bands, we can consider the weak SC pairing regime, in which case only the states around each Fermi surface will be effectively paired up. Ignoring the pairing between a state around the Fermi surface and that from other bands does not affect the topology of the system. In this way, the BdG $H'$ further reduces to an effective one band Hamiltonian projected to the middle
band. For $\mathcal{\vec{F}}_{i}$, the effective BdG Hamiltonian takes the form
\[
h_{\mathbf{Q}_{i}}(\mathbf{k},\phi)=\left[\begin{array}{cc}
\epsilon_{\mathbf{Q}_{i}+\mathbf{k}} & \Delta_{\mathbf{Q}_{i}}(\mathbf{k})\Omega_{\mathbf{Q}_{i}}(\mathbf{k})e^{-\text{i}n_{i}\phi}\\
\Delta_{\mathbf{Q}_{i}}^{*}(\mathbf{k})\Omega_{\mathbf{Q}_{i}}(\mathbf{k})e^{\text{i}n_{i}\phi} & -\epsilon_{\mathbf{Q}_{i}-\mathbf{k}}
\end{array}\right],
\]
where $\Delta_{\mathbf{Q}_{i}}\equiv\langle u_{\mathbf{k}}|\hat{\Delta}_{\mathbf{Q}_{i}}|u^{*}_{-\mathbf{k}}\rangle$ is the superconductor order parameter
projected onto the $i$-th Fermi surface. Furthermore, in each $\mathcal{\vec{F}}_{i}$,
only one $\Delta_{\mathbf{Q}_{i}}$ is non-vanishing and captures
the Berry curvature in the corresponding $\mathcal{\vec{F}}_{i}$
only, so the Chern-Simons invariant $\nu_{3}$ can be written as a
sum of the corresponding ``Chern-Simons invariant'' in different
$\mathcal{\vec{F}}_{i}$. Since $\mathcal{A}_{k_{x},k_{y},\phi}$
are local functions of $(k_{x},k_{y},\phi)$, and for effective only
one band, $\mathcal{A}_{k_{x},k_{y},\phi}$ are just numbers, hence
$\mathcal{A}^{3}=0$, thus we can decompose $\nu_{3}$ as
\[
\nu_{3}=-\frac{1}{4\pi^{2}}\sum_{i}\int_{\mathcal{\vec{F}}_{i}\times S^{1}}\mathcal{Q}_{3}[h_{\mathbf{Q}_{i}}(\mathbf{k},\phi)]=-\frac{1}{4\pi^{2}}\sum_{i}\int_{\mathcal{\vec{F}}_{i}\times S^{1}}\mathcal{A}d\mathcal{A}[h_{\mathbf{Q}_{i}}].
\]
In this case, the Chern-Simons invariant reduces to the Hopf invariant
that captures the linking number of the inverse images of two points
in the target space $S^{2}$ of $h_{\mathbf{Q}_{i}}(\mathbf{k},\phi)$.

For brevity, we consider below a particular $\mathcal{\vec{F}}=\mathcal{\vec{F}}_{i}$
for some $i$ and ignore the subscript $i$. To proceed, one needs the eigenvectors of $h_{\mathbf{Q}}(\mathbf{k},\phi)$.
In the original basis, the eigenvectors for the middle band are of
the form $|\psi_{\mathbf{k}\pm}\rangle=(\alpha_{\mathbf{k}\pm}u_{\mathbf{k}},\beta_{\mathbf{k}\pm}u_{-\mathbf{k}}^{*})^{\text{T}}$,
where $(\alpha_{\mathbf{k}\pm},\beta_{\mathbf{k}\pm})^{\text{T}}$
are the eigenvectors of $h_{\mathbf{Q}}(\mathbf{k},\phi)$ written in the
eigen-basis. For $h_{\mathbf{Q}}(\mathbf{k},\phi)$, there are nonetheless
two choices of eigenvectors (without normalization),
\begin{equation}
\left(\begin{array}{c}
\alpha_{+,\mathbf{k}\pm}\\
\beta_{+,\mathbf{k}\pm}
\end{array}\right)\propto\left(\begin{array}{c}
\xi_{\mathbf{k}}\pm\sqrt{\xi_{\mathbf{k}}^{2}+|\Delta_{\mathbf{Q}}(\mathbf{k})\Omega_{\mathbf{Q}}(\mathbf{k})|^{2}}\\
\Delta_{\mathbf{Q}}^{*}(\mathbf{k})\Omega_{\mathbf{Q}}(\mathbf{k})e^{\text{i}n\phi}
\end{array}\right)\label{eq:hole}
\end{equation}
or
\begin{equation}
\left(\begin{array}{c}
\alpha_{-,\mathbf{k}\pm}\\
\beta_{-,\mathbf{k}\pm}
\end{array}\right)\propto\left(\begin{array}{c}
\Delta_{\mathbf{Q}}(\mathbf{k})\Omega_{\mathbf{Q}}(\mathbf{k})e^{-\text{i}n\phi}\\
-\xi_{\mathbf{k}}\pm\sqrt{\xi_{\mathbf{k}}^{2}+|\Delta_{\mathbf{Q}}(\mathbf{k})\Omega_{\mathbf{Q}}(\mathbf{k})|^{2}}
\end{array}\right),\label{eq:electron}
\end{equation}
where $\xi_{\mathbf{k}}\equiv\frac{\epsilon_{\mathbf{Q}+\mathbf{k}}+\epsilon_{\mathbf{Q}-\mathbf{k}}}{2}$.
One can check that both of them are the eigenvectors of $h_{\mathbf{Q}}(\mathbf{k},\phi)$.
Actually, for Fermi surfaces, there are two cases with $\xi_{\mathbf{k}}>0$
or $\xi_{\mathbf{k}}<0$ inside the region $\vec{S}$. For these two
different cases, we choose different eigenvector, i.e., Eq. \eqref{eq:hole}
for $\xi_{\mathbf{k}}>0$ and Eq. \eqref{eq:electron} for $\xi_{\mathbf{k}}<0$.

The Berry connections are $\mathcal{A}_{\eta}=\text{i}\langle\psi_{\mathbf{k}-}|\partial_{\eta}|\psi_{\mathbf{k}-}\rangle$,
with $\eta=k_{x},k_{y},\phi$, or explicitly

\begin{align*}
\mathcal{A}_{\pm,\phi} & =\text{i}(\alpha_{\pm,\mathbf{k}-}^{*}u_{\mathbf{k}}^{\dagger},\beta_{\pm,\mathbf{k}-}^{*}u_{-\mathbf{k}}^{\text{T}})\partial_{\phi}\left(\begin{array}{c}
\alpha_{\pm,\mathbf{k}-}u_{\mathbf{k}}\\
\beta_{\pm,\mathbf{k}-}u_{-\mathbf{k}}^{*}
\end{array}\right)\\
 & =\begin{cases}
\text{i}\beta_{+,\mathbf{k}-}^{*}\partial_{\phi}\beta_{+,\mathbf{k}-}=-n|\beta_{+,\mathbf{k}-}|^{2}\text{\ensuremath{\xrightarrow{\Delta_{\mathbf{Q}}\rightarrow0}}}-n\Theta_{\vec{S}}, & \text{for }\xi_{\mathbf{k}}>0\text{ in }\vec{S}\\
\text{i}\alpha_{-,\mathbf{k}-}^{*}\partial_{\phi}\alpha_{-,\mathbf{k}-}=+n|\alpha_{-,\mathbf{k}-}|^{2}\text{\ensuremath{\xrightarrow{\Delta_{\mathbf{Q}}\rightarrow0}}}+n\Theta_{\vec{S}}, & \text{for }\xi_{\mathbf{k}}<0\text{ in }\vec{S}
\end{cases},
\end{align*}
\begin{align*}
\mathcal{A}_{\pm,\mathbf{k}} & \equiv(\mathcal{A}_{\pm,k_{x}},\mathcal{A}_{\pm,k_{y}})=\text{i}(\alpha_{\pm,\mathbf{k}-}^{*}u_{\mathbf{k}}^{\dagger},\beta_{\pm,\mathbf{k}-}^{*}u_{-\mathbf{k}}^{\text{T}})\nabla_{\mathbf{k}}\left(\begin{array}{c}
\alpha_{\pm,\mathbf{k}-}u_{\mathbf{k}}\\
\beta_{\pm,\mathbf{k}-}u_{-\mathbf{k}}^{*}
\end{array}\right)\\
 & =\text{i}(|\alpha_{\pm,\mathbf{k}-}|^{2}-|\beta_{\pm,\mathbf{k}-}|^{2})u_{\mathbf{k}}^{\dagger}\nabla_{\mathbf{k}}u_{\mathbf{k}}+\text{i}(\alpha_{\pm,\mathbf{k}-}^{*}\nabla_{\mathbf{k}}\alpha_{\pm,\mathbf{k}-}+\beta_{\pm,\mathbf{k}-}^{*}\nabla\beta_{\pm,\mathbf{k}-})\\
 & =\pm(1-2\Theta_{\vec{S}})\mathcal{A}_{0,\mathbf{k}}+\mathcal{A}_{\pm,1,\mathbf{k}},
\end{align*}
here we have used the trick $\Delta_{\mathbf{Q}}\to0$ without closing
the bulk gap, the upper (lower) sign means $\xi_{\mathbf{k}}>0$ ($\xi_{\mathbf{k}}<0$)
inside the region $\vec{S}$, $\Theta_{\vec{S}}\equiv\Theta(\xi_{\mathbf{k}}>0)$
($\Theta(\xi_{\mathbf{k}}<0)$) for the case $\xi_{\mathbf{k}}>0$
($\xi_{\mathbf{k}}<0$), and denotes the step function is 1 inside
the region $\vec{S}$ and 0 else, and $\nabla_{\mathbf{k}}\equiv(\partial_{k_{x}},\partial_{k_{y}})$.
In the last line, we denote $\mathcal{A}_{0,\mathbf{k}}\equiv\text{i}u_{\mathbf{k}}^{\dagger}\nabla_{\mathbf{k}}u_{\mathbf{k}}$,
and

\begin{align*}
\mathcal{A}_{\pm,1,\mathbf{k}} & \equiv\text{i}(\alpha_{\pm,\mathbf{k}-}^{*}\nabla_{\mathbf{k}}\alpha_{\pm,\mathbf{k}-}+\beta_{\pm,\mathbf{k}-}^{*}\nabla_{\mathbf{k}}\beta_{\pm,\mathbf{k}-})\\
 & =\pm\text{i}\frac{\Delta_{\mathbf{Q}}\Omega_{\mathbf{Q}}\nabla_{\mathbf{k}}(\Delta_{\mathbf{Q}}^{*}\Omega_{\mathbf{Q}})-\Delta_{\mathbf{Q}}^{*}\Omega_{\mathbf{Q}}\nabla_{\mathbf{k}}(\Delta_{\mathbf{Q}}\Omega_{\mathbf{Q}})}{4\sqrt{\xi_{\mathbf{k}}^{2}+\Omega_{\mathbf{Q}}^{2}|\Delta_{\mathbf{Q}}|^{2}}\left(\sqrt{\xi_{\mathbf{k}}^{2}+\Omega_{\mathbf{Q}}^{2}|\Delta_{\mathbf{Q}}|^{2}}\mp\xi_{\mathbf{k}}\right)}\\
 & \text{\ensuremath{\xrightarrow{\Delta_{\mathbf{Q}}\rightarrow0}\pm}}\text{i}\Theta_{\vec{S}}\frac{\Delta_{\mathbf{Q}}\Omega_{\mathbf{Q}}\nabla_{\mathbf{k}}(\Delta_{\mathbf{Q}}^{*}\Omega_{\mathbf{Q}})-\Delta_{\mathbf{Q}}^{*}\Omega_{\mathbf{Q}}\nabla_{\mathbf{k}}(\Delta_{\mathbf{Q}}\Omega_{\mathbf{Q}})}{2|\Delta_{\mathbf{Q}}\Omega_{\mathbf{Q}}|^{2}}.
\end{align*}

Let's denote $\vec{v}=\text{i}\frac{\Delta_{\mathbf{Q}}\Omega_{\mathbf{Q}}\nabla_{\mathbf{k}}(\Delta_{\mathbf{Q}}^{*}\Omega_{\mathbf{Q}})-\Delta_{\mathbf{Q}}^{*}\Omega_{\mathbf{Q}}\nabla_{\mathbf{k}}(\Delta_{\mathbf{Q}}\Omega_{\mathbf{Q}})}{2|\Delta_{\mathbf{Q}}\Omega_{\mathbf{Q}}|^{2}}$,
direct substitution of $\Delta_{\mathbf{Q}}=\Delta_{\mathbf{k}}e^{\text{i}\theta_{\mathbf{k}}}$
seems to give $\vec{v}=\nabla_{\mathbf{k}}\theta_{\mathbf{k}}$. But
note that $\theta_{\mathbf{k}}$ is a multivalued field, for a general
SC order parameter with phase winding $m$, there will be a branch
cut from $2\pi m\to0$. Consider the region close to the branch cut,
$\theta_{\mathbf{k}}$ behaves like $2\pi m\Theta$, where $\Theta$
is a step function, so the correct result should be \cite{Kleinert2008}
\[
\vec{v}=\nabla_{\mathbf{k}}\theta_{\mathbf{k}}+\mathbf{A}_{d},
\]
where the vector field $\mathbf{A}_{d}$ is the \emph{defect gauge
field} compensating for the discontinuity in $\theta_{\mathbf{k}}$
and the properties are
\begin{align}
\nabla_{\mathbf{k}}\times\nabla_{\mathbf{\mathbf{k}}}\theta_{\mathbf{k}} & =0,\label{eq:defect1}\\
\oint_{\partial\vec{S}}\nabla_{\mathbf{k}}\theta_{\mathbf{k}}\cdot d\mathbf{k} & =2\pi m,\label{eq:defect2}\\
\nabla_{\mathbf{k}}\times\mathbf{A}_{d} & =2\pi m\delta(\mathbf{k})\hat{z}=\hat{z}\delta(\mathbf{k})\oint_{\partial\vec{S}}\nabla_{\mathbf{\mathbf{k}}}\theta_{\mathbf{k}}\cdot d\mathbf{k},\label{eq:defect3}\\
\oint_{\partial\vec{S}}\mathbf{A}_{d}\cdot d\mathbf{k} & =\begin{cases}
0 & \text{if }\partial\text{ avoids the branch cut}\\
2\pi m & \text{otherwise}
\end{cases},\label{eq:defect4}
\end{align}
where $\partial\vec{S}$ is the boundary of the region containing
the origin. Eq. \eqref{eq:defect3} is corresponded to the boundary
of the branch cut. In order to make these results consistent, the
chain rule of differentiation must be modified to
\[
\nabla_{\mathbf{k}}e^{\text{i}\theta_{\mathbf{k}}}=\text{i}(\nabla_{\mathbf{k}}\theta_{\mathbf{k}}+\mathbf{A}_{d})e^{\text{i}\theta_{\mathbf{k}}}.
\]
Note that $\vec{v}$ is invariant under the gauge transformations
\begin{align*}
\theta_{\mathbf{k}} & \to\theta_{\mathbf{k}}+\lambda_{\mathbf{k}},\\
\mathbf{A}_{d} & \to\mathbf{A}_{d}-\nabla_{\mathbf{k}}\lambda_{\mathbf{k}}.
\end{align*}
An example is that, for a $p+\text{i}p$ SC, the SC order parameter
can be expressed as $\Delta=\Delta_{\mathbf{k}}e^{\text{i}\phi_{\mathbf{k}}}$,
with the azimuthal angle $\phi_{\mathbf{k}}\in[0,2\pi)$. In this
case, the field $\theta_{\mathbf{k}}=\phi_{\mathbf{k}}$ is discontinuous
over the positive $k_{x}$-axis, thus $\nabla_{\mathbf{k}}\theta_{\mathbf{k}}=\hat{\phi}/k-2\pi\Theta(k_{x})\delta(k_{y})\hat{k}_{y}$.
Note that we expect $\vec{v}=\hat{\phi}/k$, hence $\vec{v}=\nabla_{\mathbf{k}}\theta_{\mathbf{k}}+\mathbf{A}_{d}\text{, and }\mathbf{A}_{d}=2\pi\Theta(k_{x})\delta(k_{y})\hat{k}_{y}$.

Now, return to our proof for the Chern-Simons invariant. Consider
the defect gauge field, we have
\[
\mathcal{A}_{\pm,1,\mathbf{k}}=\pm\Theta_{\vec{S}}(\nabla_{\mathbf{k}}\text{arg}\Delta_{\mathbf{Q}}+\mathbf{A}_{d}).
\]
In our effective one band case, since $\mathcal{A}d\mathcal{A}=[\mathcal{A}_{k_{x}}(\partial_{k_{y}}\mathcal{A}_{\phi}-\partial_{\phi}\mathcal{A}_{k_{y}})+\mathcal{A}_{k_{y}}(\partial_{\phi}\mathcal{A}_{k_{x}}-\partial_{k_{x}}\mathcal{A}_{\phi})+\mathcal{A}_{\phi}(\partial_{k_{x}}\mathcal{A}_{k_{y}}-\partial_{k_{y}}\mathcal{A}_{k_{x}})]dk_{x}\wedge dk_{y}\wedge d\phi$,
and $\partial_{\phi}\mathcal{A}_{k_{x},k_{y}}=0$, one can readily
show that
\begin{equation*}
  \nu_{3}=\sum_{i}\nu^{(i)}_{3}\;\mathrm{mod}\;2,
\end{equation*}
and
\begin{align*}
\nu^{(i)}_{3} & =-\frac{1}{2\pi}\left(\frac{1}{2\pi}\int_{0}^{2\pi}d\phi\right)\int_{\mathcal{\vec{F}}_{i}}\left[\mathcal{A}_{i,\pm,\phi}\nabla_{\mathbf{k}}\times\mathcal{A}_{i,\pm,\mathbf{k}}+\mathcal{A}_{i,\pm,\mathbf{k}}\times\nabla_{\mathbf{k}}\mathcal{A}_{i,\pm,\phi}\right]\cdot d^{2}\mathbf{k}\\
 & =\pm n_{i}\frac{1}{2\pi}\int_{\mathcal{\vec{F}}_{i}}\left[\Theta_{\vec{S}_{i}}\nabla_{\mathbf{k}}\times\mathcal{A}_{i,\pm,\mathbf{k}}+\mathcal{A}_{i,\pm,\mathbf{k}}\times\nabla_{\mathbf{k}}\Theta_{\vec{S}_{i}}\right]\cdot d^{2}\mathbf{k}.
\end{align*}
For the part involving $\mathcal{A}_{i,0,\mathbf{k}}$, we have
\begin{align*}
\nu_{3}^{(i)[\mathcal{A}_{i,0,\mathbf{k}}]} & =\frac{n_{i}}{2\pi}\int_{\mathcal{\vec{F}}_{i}}\left[\Theta_{\vec{S}_{i}}\nabla_{\mathbf{k}}\times[(1-2\Theta_{\vec{S}_{i}})\mathcal{A}_{i,0,\mathbf{k}}]+(1-2\Theta_{\vec{S}_{i}})\mathcal{A}_{i,0,\mathbf{k}}\times\nabla_{\mathbf{k}}\Theta_{\vec{S}_{i}}\right]\cdot d^{2}\mathbf{k}\\
 & =-\frac{n_{i}}{2\pi}\int_{\mathcal{\vec{F}}_{i}}[\Theta_{\vec{S}_{i}}\nabla_{\mathbf{k}}\times\mathcal{A}_{i,0,\mathbf{k}}-\mathcal{A}_{i,0,\mathbf{k}}\times\nabla_{\mathbf{k}}\Theta_{\vec{S}_{i}}+2\Theta_{\vec{S}_{i}}\nabla_{\mathbf{k}}\Theta_{\vec{S}_{i}}\times\mathcal{A}_{i,0,\mathbf{k}}\\
  &\hspace{14pt}+2\Theta_{\vec{S}_{i}}\mathcal{A}_{i,0,\mathbf{k}}\times\nabla_{\mathbf{k}}\Theta_{\vec{S}_{i}}]\cdot d^{2}\mathbf{k}\\
 & =-\frac{n_{i}}{2\pi}\left(\int_{\vec{S}_{i}}\nabla_{\mathbf{k}}\times\mathcal{A}_{i,0,\mathbf{k}}\cdot d^{2}\mathbf{k}-\oint_{\partial\vec{S}_{i}}\mathcal{A}_{i,0,\mathbf{k}}\cdot d\mathbf{k}\right).
\end{align*}
Since the Chern-Simons invariant is gauge-independent, we can consider
a smooth gauge, using the Stokes' theorem, $\nu_{3}^{(i)[\mathcal{A}_{i,0,\mathbf{k}}]}$
just vanishes.

Next we need to consider for $\pm\Theta_{\vec{S}_{i}}(\nabla_{\mathbf{k}}\text{arg}\Delta_{\mathbf{Q}_{i}}+\mathbf{A}_{i,d})$,
which can be divided into two parts $\pm\Theta_{\vec{S}_{i}}\mathbf{A}'_{i}$,
with $\mathbf{A}'_{i}=\nabla_{\mathbf{k}}\text{arg}\Delta_{\mathbf{Q}_{i}}$
or $\mathbf{A}_{i,d}$,
\begin{align*}
\nu_{3}^{(i)[\mathbf{A}'_{i}]} & =\frac{n_{i}}{2\pi}\int_{\mathcal{\vec{F}}_{i}}\left[\Theta_{\vec{S}_{i}}\nabla_{\mathbf{k}}\times[\Theta_{\vec{S}_{i}}\mathbf{A}'_{i}]+\Theta_{\vec{S}_{i}}\mathbf{A}'_{i}\times\nabla_{\mathbf{k}}\Theta_{\vec{S}_{i}}\right]\cdot d^{2}\mathbf{k}\\
 & =\frac{n_{i}}{2\pi}\int_{\mathcal{\vec{F}}_{i}}\left[\Theta_{\vec{S}_{i}}\nabla_{\mathbf{k}}\times\mathbf{A}'_{i}+\Theta_{\vec{S}_{i}}\nabla_{\mathbf{k}}\Theta_{\vec{S}_{i}}\times\mathbf{A}'_{i}+\Theta_{\vec{S}_{i}}\mathbf{A}'_{i}\times\nabla_{\mathbf{k}}\Theta_{\vec{S}_{i}}\right]\cdot d^{2}\mathbf{k}\\
 & =\frac{n_{i}}{2\pi}\int_{\mathcal{\vec{F}}_{i}}\Theta_{\vec{S}_{i}}\nabla_{\mathbf{k}}\times\mathbf{A}'_{i}\cdot d^{2}\mathbf{k}=\frac{n_{i}}{2\pi}\int_{\vec{S}_{i}}\nabla_{\mathbf{k}}\times\mathbf{A}'_{i}\cdot d^{2}\mathbf{k}.
\end{align*}
Since $\nabla_{\mathbf{k}}\times\nabla_{\mathbf{k}}\text{arg}\Delta_{\mathbf{Q}_{i}}=0$,
$\nu_{3}^{(i)[\nabla_{\mathbf{k}}\text{arg}\Delta_{\mathbf{Q}_{i}}]}$ vanishes.
Next, by the property \eqref{eq:defect3} of $\mathbf{A}_{i,d}$,
\begin{equation*}
\nu_{3}^{(i)[\mathbf{A}_{i,d}]}=\frac{n_{i}}{2\pi}\int_{\vec{S}_{i}}\nabla_{\mathbf{k}}\times\mathbf{A}_{i,d}\cdot d^{2}\mathbf{k}=\frac{n_{i}}{2\pi}\oint_{\partial\vec{S}_{i}}\nabla_{\mathbf{k}}\text{arg}\Delta_{\mathbf{Q}_{i}}\cdot d\mathbf{k}\int_{\vec{S}_{i}}\delta(\mathbf{k})\hat{z}\cdot d^{2}\mathbf{k}=\frac{n_{i}}{2\pi}\oint_{\partial\vec{S}_{i}}\nabla_{\mathbf{k}}\text{arg}\Delta_{\mathbf{Q}_{i}}\cdot d\mathbf{k}.
\end{equation*}
Note that this is \emph{not} Stokes' theorem. Gathering
all the results together, we have
\[
\nu_{3}=\frac{1}{2\pi}\sum_{i}n_{i}\oint_{\partial\vec{S}_{i}}\nabla_{\mathbf{k}}\arg\Delta_{\mathbf{Q}_{i}}(\mathbf{k})\cdot d\mathbf{k}\;\mathrm{mod}\;2.
\]

Note that the pairing $\Delta_{\mathbf{Q}_{i}}(\mathbf{k})$ has captured
the band topology. For the case with SC pairings between two different
Fermi surfaces, using the similar arguments as in \cite{LiuX-J2017},
the formula is still applicable. This completes the proof.
\end{document}